\documentclass[letterpaper, 10 pt, conference]{ieeeconf} 

\usepackage{amsfonts,amssymb,amsmath}
\usepackage{graphicx,graphics,epsfig,color}
\usepackage{rgsMacros}
\usepackage{subcaption}
\usepackage{hyperref}

\let\labelindent\relax
\usepackage{enumitem}
\usepackage{balance}

\definecolor{olivegreen}{rgb}{0.14,0.29,0}

\IEEEoverridecommandlockouts                              
\newtheorem{exe}{Example}
\newtheorem{corol}{Corollary}
\newtheorem{ass}{Assumption}
\newtheorem{proper}{Property}
\newtheorem{defin}{Definition}
\newtheorem{prob}{Problem}
\newtheorem{cla}{Claim}
\newtheorem{rem}{Remark}
\newtheorem{lem}{Lemma}
\newtheorem{prop}{Proposition}
\newtheorem{thm}{Theorem}
\newtheorem{fct}{Fact}

\newenvironment{example}{\begin{exe}}{\hfill $\square$ \end{exe}}
\newenvironment{remark}{\begin{rem} \rm}{\hfill $ $ \end{rem}}

\newenvironment{theorem}{\begin{thm}}{\hfill $ $ \end{thm}}

\usepackage{mathrsfs}  
\usepackage{comment}

\newif\ifitsdraft

\newtheorem{dwellt}{Condition}

\newif\ifitsdraft

\usepackage{mathtools}
\definecolor{cadmiumgreen}{rgb}{0.0, 0.42, 0.24}
\usepackage{tikz}
\usetikzlibrary{shapes,positioning,intersections,quotes}
\usepackage{pgfplots}
\pgfplotsset{compat=1.16}

\tikzset{
    cross/.pic = {
    \draw[rotate = 45] (-#1,0) -- (#1,0);
    \draw[rotate = 45] (0,-#1) -- (0, #1);
    }
}
\usepackage{amsfonts,amssymb,amsmath}

\usepackage{xcolor}

\usepackage{subcaption}

\newlength{\overwritelength}
\newlength{\minimumoverwritelength}
\newcommand{\overwrite}[3][red]{%
  \settowidth{\overwritelength}{$#2$}%
  \ifdim\overwritelength<\minimumoverwritelength%
    \setlength{\overwritelength}{\minimumoverwritelength}\fi%
  \stackrel
    {%
      \begin{minipage}{\overwritelength}%
        \color{#1}\centering\small #3\\%
        \rule{1pt}{9pt}%
      \end{minipage}}
    {\colorbox{#1!50}{\color{black}$\displaystyle#2$}}}

\usepackage{color}

\usepackage{pifont}
\usetikzlibrary{arrows.meta,positioning,calc,decorations.pathmorphing}

\tikzset{
  snakeline/.style = {->,thick, decorate, decoration={pre length=0.2cm, post length=0.2cm, snake, amplitude=.4mm, segment length=2mm}, cadmiumgreen},
  block/.style = {draw, fill=blue!20, minimum height=3em, minimum width=3em},
  pinstyle/.style={pin edge={to-,thin,black}},
}

\usetikzlibrary{shapes.misc}

\tikzset{cross/.style={cross out, draw=black, minimum size=2*(#1-\pgflinewidth), inner sep=0pt, outer sep=0pt},
cross/.default={1pt}}

\makeatletter
\newcommand \EmptyArgParse [2]
  {%
    \if\relax\detokenize{#2}\relax
      \def\ProcessedArgument{#1}%
    \else
      \def\ProcessedArgument{#2}%
    \fi
  }
\NewDocumentCommand \mathcircled { >{\EmptyArgParse{red}}O{} O{circle} m }
  {%
    \mathpalette{\mathcircled@b{#1}{#2}}{#3}%
  }
\newcommand\mathcircled@b[4]
  {%
    \tikz[baseline=(math.base)]
      \node[draw,color=#1,#2,inner sep=1pt] (math) {$\m@th#3#4$};%
  }
\makeatother

\title{\LARGE \bf Inverse Optimal Cardano-Lyapunov Feedback\\ for PDEs with Convection}

\author{M. C. Belhadjoudja, M. Krsti{\'c}, M. Maghenem, and E. Witrant 
\thanks{M. C. Belhadjoudja and M. Maghenem are with Universit\'e Grenoble Alpes, CNRS, Grenoble-INP, GIPSA-lab, F-38000, Grenoble, France (e-mail: mohamed.belhadjoudja@gipsa-lab.fr).}
\thanks{E. Witrant is with Universit\'e Grenoble Alpes, CNRS, Grenoble-INP, GIPSA-lab, F-38000, Grenoble, France, and the Departement of Mechanical Engineering, Dalhousie University, Halifax B3H 4R2, Nova Scotia, Canada.}
\thanks{
M. Krsti{\'c} is with the Department of Mechanical and Aerospace Engineering, University of California San Diego, 92093
San Diego, USA (e-mail: krstic@ucsd.edu).}
\thanks{The project has been partially supported by CNRS, \textit{appel unique}, 2022.}}

\begin{document}

\maketitle

\begin{abstract}
We consider the problem of inverse optimal control design for systems that are not affine in the control. In particular, we consider some classes of partial differential equations (PDEs) with quadratic convection and counter-convection, for which the $L^2$ norm is a control Lyapunov function (CLF) whose derivative has either a depressed cubic or a quadratic dependence in the boundary control input. We also consider diffusive PDEs with or without linear convection, for which a weighted $L^2$ norm is a CLF whose derivative has a quadratic dependence in the control input. For each structure on the derivative of the CLF, we achieve inverse optimality with respect to a meaningful cost functional. For the case where the derivative of the CLF has a depressed cubic dependence in the control, we construct a cost functional for which the unique minimizer is the unique real root of a cubic polynomial: the Cardano-Lyapunov controller. When the derivative of the CLF is quadratic in the control, we construct a cost functional that is minimized by two distinct feedback laws, that correspond to the two distinct real roots of a quadratic equation. We show how to switch from one root to the other to reduce the control effort. 
\end{abstract}

\section{Introduction and Main Results} \label{intro+main_results}

Parabolic PDEs with convection arise in fluids, traffic, chemical reactors, manufacturing, and other applications. They give rise to a fascinating feature that, under Dirichlet boundary actuation, the simple $L^2$ spatial norm is a control Lyapunov function (CLF) due to the fact that the CLF's derivative is not affine in control but a higher order polynomial in control with a non-zero leading coefficient. In \cite{cardano_lyapu} we introduced feedback laws for such systems, which achieve stabilization far more simply than using PDE backstepping. In this paper, we expand that family of stabilizing controllers to controllers that are inverse optimal with respect to meaningful cost functionals, which are positive definite but not quadratic in control. 

\subsection{Problem Statement and Prior Work}
Take a control system (finite or infinite dimensional), with state $u$ and scalar control input $v$. Suppose that this system admits the origin $(u,v) := (0,0)$ as an equilibrium point. Moreover, suppose that we know a CLF $V$ for this system whose time derivative along the control system has the form 
\begin{align}
\dot{V}(u) = \phi(u) + \beta (u)v + v^3,\label{depressed_cubic}
\end{align}
where $\phi$ and $\beta$ are continuous operators verifying $\phi(0)=\beta(0)=0$. A typical example of control systems admitting a CLF with the depressed cubic structure in \eqref{depressed_cubic} is given by the class of parabolic PDEs with quadratic convection, in which case $V$ is simply the $L^2$ norm. Very few works highlight this fundamental property of Dirichlet boundary-actuated convective parabolic PDEs, such as \cite[Equation 3.2]{burgerskristic} for Burgers' equation, and \cite[Equation 4.19]{KS_krstic}, \cite[Lemma 6]{KS1}, \cite[Lemma 3]{ACC23-KS}, and \cite[Equations 11 and 12]{auto_KS} for the Kuramoto-Sivashinsky equation. For instance, for the generalized Burgers' equation, the functions $\phi$ and $\beta$ are given explicitly below. 
\begin{example}[PDEs with quadratic convection]\label{ex1}
Consider the class of PDEs 
\begin{align}
u_{t} = \epsilon u_{xx} - (u^2)_{x} + \mathcal{R}(u) \quad x\in (0,1) \label{burgers_conv1}
\end{align}
subject to Dirichlet-type actuation 
\begin{align}
u(0) = v, \ \ u(1) = 0, \label{eq2-2}
\end{align}
where $u$ is the state and $v$ is the control input. Here, $\epsilon >0$ is the diffusion coefficient, and $\mathcal{R}$ is a continuous operator such that $\mathcal{R}(0)=0$. In this particular case, the function $V(u) := \frac{3}{4}\int_{0}^{1}u^2dx$ is a CLF and $\dot{V}$ is of the form \eqref{depressed_cubic}, with 
\begin{align}
\phi := \frac{3}{2}\left(\int_{0}^{1}u\mathcal{R}(u)dx-\epsilon \int_{0}^{1}u_{x}^2dx\right),\ \ \beta := -\frac{3\epsilon }{2}u_{x}(0). \nonumber
\end{align}
\end{example}
Let $\alpha$ be a class $\mathcal{K}_{\infty}$ function, and suppose that we want to design $v$ such that along the closed-loop solutions, we have 
\begin{align}
\dot{V} \leq -\alpha (V). \label{exp_stab}
\end{align}
E.g., choosing $\alpha (V):=V$, \eqref{exp_stab} would imply exponential stability. This stabilization problem is addressed in \cite{cardano_lyapu}, inspired by the ideas developed in \cite{burgerskristic,KS_krstic,KS1,ACC23-KS,auto_KS}. We set the feedback law as
\begin{align}
v := \kappa_{c}(\phi (u), \beta (u), V(u)), \label{feedback}
\end{align}
that is continuous with respect to its arguments, vanishes at the origin $u:=0$, and leads to asymptotic stability in the sense of inequality \eqref{exp_stab}. This feedback law is given by the Cardano-Lyapunov formula 
\begin{equation}
\kappa_{c}(u) := \sqrt[3]{-\frac{q}{2}+\sqrt{\frac{q^2}{4}+ \frac{\beta^3}{27}}}+\sqrt[3]{-\frac{q}{2}-\sqrt{\frac{q^2}{4}+ \frac{\beta^3}{27}}},\label{cardano_lyapu1}
\end{equation}
where 
\begin{align}
q(u):= |\phi(u)|+\frac{2\sqrt{3}}{9}|\beta|^{\frac{3}{2}}+\alpha (V).\label{cardano_lyapu2}
\end{align}
By injecting \eqref{cardano_lyapu1}-\eqref{cardano_lyapu2} in \eqref{depressed_cubic}, we find 
\begin{align}
\dot{V} = \phi - |\phi|-\frac{2\sqrt{3}}{9}|\beta|^{\frac{3}{2}}-\alpha (V),
\end{align}
which implies \eqref{exp_stab}. 

The question we would like to answer in this paper is whether this feedback law is optimal, for some \textit{meaningful} cost functional. This defines the concept of \textit{inverse optimality}. Namely, instead of choosing a cost functional and deriving a corresponding optimal feedback law, we construct a (family of) feedback law(s) and show that some cost functional is minimized in closed-loop. We say in this case that the (family of) control law(s) is \textit{inverse optimal}. 

The origin of the inverse optimal control approach goes back to the works of Kalman on robustness analysis of linear quadratic regulators \cite{kalman}. The first extension of the inverse optimal control concept to nonlinear control systems affine in the control is due to Moylan and Anderson in \cite{moylan_anderson}, and a complete methodology for the design of robust inverse optimal nonlinear controllers for control-affine systems has been developed in \cite{freeman_kokotovic,constructive}. Since then, a vast amount of work has been devoted to inverse optimal control, and it would not be possible to survey all the obtained results here comprehensively. To cite just few references, in \cite{adaptive_inverse},
adaptive inverse optimal controllers are designed for systems that are affine in the control input and in the unknown parameters. The cost that is minimized involves penalty on the state, the control, and the adaptation error. In \cite{Krstic_inverse_stochastic1}, inverse optimal controllers are designed for stochastic nonlinear systems that are affine in the control and in the noise. The cost is minimized by the control and maximized by the noise. In \cite{Krstic_ISS}, nonlinear systems affine in the control and in a deterministic disturbance are considered, and inverse optimal controllers, achieving input-to-state stability with respect to the disturbance, are constructed. The cost is minimized by the controller and maximized by the disturbance. In \cite{stochastic2}, inverse optimal controllers are designed for stochastic nonlinear systems that are affine in the control and in the noise, under the assumption that the covariance of the noise is unknown. More recently, the inverse optimal control concept has been successfully adapted to the problem of inverse optimal safety filter design for nonlinear systems that are affine in the control \cite{Krstic_safety}. 

The common point in the aforementioned references is that the underlying control system is affine in the control. In this case, the inverse optimal controllers can be designed using Sontag's universal formula \cite{sontag_universal}. This approach is specific to control-affine systems, and cannot be applied if the derivative of the CLF is given, e.g., by \eqref{depressed_cubic}. Another situation where the existing results developed so far cannot be applied is when the derivative of the CLF $V$ is of the form 
\begin{align}
\dot{V}(u) = \phi (u) + \beta (u)v - v^2. \label{quadratic_vdot}
\end{align}
It has been shown in \cite[Equation 13]{espitia_lyapunov} that linear non-convective diffusion-reaction equations subject to a right-end Dirichlet-type actuation admit a weighted $L^2$ norm as a CLF, whose derivative has the quadratic structure in \eqref{quadratic_vdot}. Another example of a control system admitting a CLF whose derivative has the structure in \eqref{quadratic_vdot} is given below.  
\begin{example}[PDEs with counter-convection]\label{ex2}
Consider the equation
\begin{align}
u_{t} = \epsilon u_{xx} + u_{x} + \mathcal{R}(u) \quad x\in (0,1)\label{not_burgers}
\end{align}
subject to Dirichlet actuation \eqref{eq2-2}. In this case, the function $V(u) := \int_{0}^{1}u^2dx$ is a CLF whose derivative verifies \eqref{quadratic_vdot} with 
\begin{align}
&\phi :=2\left(\int_{0}^{1}u\mathcal{R}(u)dx - \epsilon \int_{0}^{1}u_{x}^2dx\right),\ \ \beta := -2\epsilon u_{x}(0). \nonumber 
\end{align}
\end{example}
The structure of the feedback law developed in \cite[Equation 25]{espitia_lyapunov} for the specific case of linear diffusion-reaction PDEs is, in fact, general, and can be extended to any control system admitting a CLF whose derivative has the quadratic structure in \eqref{quadratic_vdot}, especially to some classes of boundary actuated convective PDEs; see for instance \cite{cardano_lyapu}. Namely, asymptotic stabilization, in the sense of inequality \eqref{exp_stab}, can be achieved by setting
\begin{align}
v := \kappa_{q}(\phi (u),\beta (u),V(u)), \label{feedback_quadratic} 
\end{align}
where $\kappa_{q}$ is continuous in its arguments, vanishes if $u:=0$, and is given by the formula 
\begin{align}
\kappa_{q}(u) := \frac{\beta + \sqrt{\beta^2+4\theta (u)}}{2}, \label{quadratic_feedback}
\end{align}
where 
\begin{align}
\theta (u) := |\phi (u)|+\alpha (V).\label{theta_def}
\end{align}
By injecting \eqref{quadratic_feedback} in \eqref{quadratic_vdot}, we find 
\begin{align}
\dot{V} = \phi - |\phi| - \alpha (V), \label{diff_eq_V}
\end{align}
which implies \eqref{exp_stab}. We show in this paper that, as for the Cardano-Lyapunov formula, this quadratic feedback law is inverse optimal for some meaningful cost functional. 

\subsection{Main Results}
We prove in this paper the following results.  
\begin{theorem}\label{thm1}
Consider a control system (finite or infinite dimensional) with state $u$, and scalar control input $v$. Suppose that the origin $(u,v):=(0,0)$ is an equilibrium point of the system and that we know a CLF $V$ whose derivative along the control system has the depressed cubic structure in \eqref{depressed_cubic}, for some continuous operators $\phi$ and $\beta$ that vanish at $u:=0$. Moreover, let
\begin{equation}
\displaystyle \kappa_{c}^{*}(u) := m\kappa_{c}(\phi(u), \beta(u)/m^2,V(u)) \quad m\geq 2,    \label{kcstar}
\end{equation}
where $\kappa_{c}$ is given in \eqref{feedback}, \eqref{cardano_lyapu1}, \eqref{cardano_lyapu2}. Then the feedback law
\begin{align}
v := \kappa_{c}^{*}(u)
\end{align} 
achieves asymptotic stabilization in the sense of inequality \eqref{exp_stab}, and is the unique asymptotically stabilizing minimizer of the cost functional
\begin{align}
\mathcal{J}_{c} := \int_{0}^{\infty}\left(L_{c}+R_{c} (\beta + v^2)^2 v^2\right)dt,\label{costJc}
\end{align}
where 
\begin{align}
L_{c}(u) :=&~ m(m-2)R_{c}^{-1} -2m(\phi-R_{c}^{-1})\, \geq 0\label{lc},\\[5pt]
R_{c}(u)^{-1} :=&~ \displaystyle m^2\bigg[|\phi|+\frac{2\sqrt{3}}{9}\left|\frac{\beta}{m^2}\right|^{\frac{3}{2}}+\alpha (V)\bigg] \,\geq 0.\label{rc}
\end{align}
The minimum of $\mathcal{J}_{c}$ is $\mathcal{J}_{c}^{*} := 2mV(0)$.
\end{theorem}
\begin{theorem}\label{thm2}
Consider a control system (finite or infinite dimensional) with state $u$, and scalar control input $v$. Suppose that the origin $(u,v):=(0,0)$ is an equilibrium point of the system and that we know a CLF $V$ whose derivative along the control system has the quadratic structure in \eqref{quadratic_vdot}, for some continuous operators $\phi$ and $\beta$ that vanish at $u:=0$. Moreover, let
\begin{align}
\displaystyle \kappa_{q}^{*}(u) := m\kappa_{q}(\phi(u),\beta(u)/m,V(u)) \quad m\geq 2,    \label{kq}
\end{align}
where $\kappa_{q}$ is given in \eqref{feedback_quadratic}, \eqref{quadratic_feedback}, \eqref{theta_def}. Then both feedback laws
\begin{align}
&v := \kappa_{q}^{*}(u), \ \ \text{and} \\
&v := \beta(u) - \kappa_{q}^{*}(u),
\end{align}
guarantee \eqref{exp_stab}, and minimize the cost functional 
\begin{align}
\mathcal{J}_{q} := \int_{0}^{\infty}\left(L_{q}+R_{q}\left(\beta- v\right)^2v^2\right)dt,\label{costJq}
\end{align}
where 
\begin{align}
&L_{q}(u) := -2m(\phi-R_{q}^{-1})+m(m-2)R_{q}^{-1},\label{lq}\\
&R_{q}(u)^{-1} := m\theta (u). \label{rq}
\end{align}
The minimum of $\mathcal{J}_{q}$ for both feedback laws is $\mathcal{J}_{q}^{*} := 2mV(0)$.
\end{theorem}
\begin{remark}
    Note that the functionals $L_{c}$ and $R_{c}$ defined respectively in \eqref{lc} and \eqref{rc}, verify 
    \begin{eqnarray}
     &L_c(0) =0\,, \quad R_{c}(0)= +\infty  & \\
    & \mbox{$L_c(u)>0\,, \ \ 0<R_{c}(u)< +\infty$ \ \ when \ \ $u\neq 0$.} &
\end{eqnarray}
Similarly, the functionals $L_{q}$ and $R_{q}$ defined respectively in \eqref{lq} and \eqref{rq}, verify
\begin{eqnarray}
     &L_q(0) =0\,, \quad R_{q}(0)=+\infty  & \\
    & \mbox{$L_q(u)>0\,, \ \ 0<R_{q}(u)<+\infty$ \ \ when \ \ $u\neq 0$.} &
\end{eqnarray}
\end{remark}

\begin{remark}
The cost functionals $\mathcal{J}_{c}$ and $\mathcal{J}_{q}$ are meaningful for the following reasons:
\begin{enumerate}[label={
\arabic*.},leftmargin=*]
    \item Any meaningful cost functional has to penalize the state $u$ and the control input $v$, i.e. the functionals $(u,v)\mapsto \mathscr{L}_{c}(u,v)\in \bar{\mathbb{R}} := [-\infty, \infty]$ and $(u,v)\mapsto \mathscr{L}_{q}(u,v)\in \bar{\mathbb{R}}$ defined by 
\begin{align}
&\mathscr{L}_{c} := L_{c}+R_{c}(\beta + v^2)^2v^2, \\
&\mathscr{L}_{q} := L_{q}+R_{q}(\beta - v)^2v^2,
\end{align}
need to be positive definite. This condition is verified. Indeed, if $(u,v)\neq (0,0)$, then $R_{c}>0$, and since $m\geq 2$, then $m(m-2)R_{c}^{-1}\geq 0$. Additionally, since $\alpha \in \mathcal{K}_{\infty}$, then 
\begin{align}
-2m(\phi - &R_{c}^{-1}) = -2m(\phi - m^2|\phi|) \nonumber \\
&~+2m^3\bigg(\frac{2\sqrt{3}}{9}\left|\frac{\beta}{m^2}\right|^{\frac{3}{2}}+\alpha (V)\bigg) > 0, 
\end{align}
which implies that $L_c>0$, and thus, that $\mathscr{L}_{c}$ is positive as $R_c(\beta + v^2)^2v^2 \geq 0$. Moreover, $\mathscr{L}_{c}(0,0)=0$. Similarly, if $(u,v)\neq (0,0)$, then $R_{q}>0$, and since $m\geq 2$, then $m(m-2)R_{q}^{-1}\geq 0$. Additionally, since $\alpha \in \mathcal{K}_{\infty}$ then 
\begin{align}
-2m(\phi - &R_{q}^{-1}) = -2m(\phi - m|\phi|)+2m^2\alpha (V) > 0, 
\end{align}
which implies that $L_q>0$, and thus, that $\mathscr{L}_{q}$ is positive as $R_q(\beta-v)^2v^2\geq 0$. Finally, $\mathscr{L}_{q}(0,0)=0$.  
\item It is also important to guarantee that $\mathscr{L}_{c}(0,v)>0$ and $\mathscr{L}_{q}(0,v)>0$ for all $v\neq 0$. This property is verified. Indeed, when $u=0$ and $v\neq 0$, having $\phi (0)=\beta (0)=0$, $\alpha$ a class $\mathcal{K}_{\infty}$ function, and $R_{c}(0)=R_{q}(0)=+\infty$, we conclude that the functionals $\mathscr{L}_{c}$ and $\mathscr{L}_{q}$ satisfy
\begin{align}
&\mathscr{L}_{c}(0,v) = +\infty v^6 >0,\ \ \mathscr{L}_{q}(0,v) = +\infty v^4 >0.
\end{align}
It means that for zero state, the penalty on control is infinite. This property comes from the weight on the control being inversely proportional to the state. It is interesting to observe also that $\mathscr{L}_{c}(u,0)>0$ and $\mathscr{L}_{q}(u,0)>0$ for all $u\neq 0$. 
\item Finally, note that when the state grows to $+\infty$ in the $L^2$ norm, then, since $\alpha$ is a class $\mathcal{K}_{\infty}$ function, the function $R_{c}$ that multiplies $v^2$ in $\mathcal{J}_{c}$, as well as the function $R_{q}$ that multiplies $v^2$ in $\mathcal{J}_{q}$, go to zero. It means that the optimal controller is allowed to take larger values if the state is farther from the equilibrium, which is a common property of inverse optimal feedback controllers \cite{constructive,Krstic_inverse_stochastic1,Krstic_ISS,Krstic1999inverse}.   
\end{enumerate}
\end{remark}

\subsection{Examples}
We give here some examples to illustrate the results in Theorems \ref{thm1} and \ref{thm2}. 
\begin{example}[PDEs with quadratic convection]
Consider the class of PDEs \eqref{burgers_conv1} subject to \eqref{eq2-2}, and let $V(u):=\frac{3}{4}\int_{0}^{1}u^2dx$. The derivative of $V$ has the depressed cubic structure in \eqref{depressed_cubic}, with $\phi$ and $\beta$ given in Example \ref{ex1}. According to Theorem \ref{thm1}, inverse optimal asymptotic stabilization can be achieved by setting $v:=\kappa_{c}^{*}$, where $\kappa_{c}^{*}$ is given by \eqref{kcstar}, \eqref{feedback}, \eqref{cardano_lyapu1}, \eqref{cardano_lyapu2}. The cost that is minimized, $\mathcal{J}_{c}$, is defined in \eqref{costJc}, where the functionals $L_{c}$ and $R_{c}$ are given by
\begin{align}
L_{c} :=&~ -2m\left(\frac{3}{2}\left(\int_{0}^{1}u\mathcal{R}(u)dx-\epsilon \int_{0}^{1}u_{x}^2dx\right)-R_{c}^{-1}\right) \nonumber \\
&~+m(m-2)R_{c}^{-1}, \\
R_{c}^{-1} :=&~ m^2\bigg[\bigg|\frac{3}{2}\left(\int_{0}^{1}u\mathcal{R}(u)dx-\epsilon \int_{0}^{1}u_{x}^2dx\right)\bigg| \nonumber \\
&~+\frac{\sqrt{2}\epsilon^{\frac{3}{2}}}{2}\left|\frac{u_{x}(0)}{m^2}\right|^{\frac{3}{2}}+\alpha (V)\bigg].
\end{align}
\end{example}
\begin{example}[PDEs with linear convection]\label{ex3}
Consider 
\begin{align}
u_{t} = \epsilon u_{xx} - u_{x} + \mathcal{R}(u) \quad x\in (0,1),\label{convective_pde}
\end{align}
subject to the left-end Dirichlet-type actuation \eqref{eq2-2}. The $L^2$ norm is not a CLF for this PDE. Indeed, 
\begin{align}
\frac{d}{dt}\int_{0}^{1}u^2dx =&~ 2\left(\int_{0}^{1}u\mathcal{R}(u)dx - \epsilon \int_{0}^{1}u_{x}^2dx\right)\nonumber \\
&~- 2\epsilon u_{x}(0)v + v^2. 
\end{align}
Inspired by the approach developed in \cite{espitia_lyapunov} for diffusive PDEs without convection, we suggest to consider the weighted $L^2$ norm
\begin{align}
V(u) := \int_{0}^{1}e^{-\frac{2}{\epsilon} x}u^2dx.\label{vweighted}
\end{align}
By differentiating $V$ along the regular solutions to \eqref{convective_pde} subject to \eqref{eq2-2}, and using integration by parts, we find 
\begin{align}
\dot{V} =&~ \frac{2}{\epsilon} V + 2\int_{0}^{1}e^{-\frac{2}{\epsilon}x}u(x)\mathcal{R}(u)dx \nonumber \\
&~-2\epsilon \int_{0}^{1}e^{-\frac{2}{\epsilon}x}u_{x}(x)^2dx - 2\epsilon u_{x}(0)v - v^2. 
\end{align}
The function $V$ is therefore a CLF whose derivative has the quadratic structure in \eqref{quadratic_vdot}, with $\beta(u) :=-2\epsilon u_x(0)$ and 
\begin{align}
\phi (u) :=&~ \frac{2}{\epsilon} V + 2\int_{0}^{1}e^{-\frac{2}{\epsilon}x}u(x)\mathcal{R}(u)dx \nonumber \\
&~-2\epsilon \int_{0}^{1}e^{-\frac{2}{\epsilon}x}u_{x}(x)^2dx. 
\end{align}
To achieve inverse optimal asymptotic stabilization, it is therefore sufficient to set $v := \kappa_{q}^{*}$ or $v:=\beta - \kappa_{q}^{*}$, where $\kappa_{q}^{*}$ is given by \eqref{kq}, \eqref{feedback_quadratic}, \eqref{quadratic_feedback}, \eqref{theta_def}. The cost that is minimized, $\mathcal{J}_{q}$, is defined in \eqref{costJq} with 
\begin{align}
L_{q} :=&~ -2m\bigg(\frac{2}{\epsilon} V + 2\int_{0}^{1}e^{-\frac{2}{\epsilon}x}u\mathcal{R}(u)dx \nonumber \\
&~-2\epsilon \int_{0}^{1}e^{-\frac{2}{\epsilon}x}u_{x}^2dx-R_{q}^{-1}\bigg)+m(m-2)R_{q}^{-1}, \\
R_{q}^{-1} :=&~ m^2\bigg[\bigg|\frac{2}{\epsilon} V + 2\int_{0}^{1}e^{-\frac{2}{\epsilon}x}u\mathcal{R}(u)dx \nonumber \\
&~-2\epsilon \int_{0}^{1}e^{-\frac{2}{\epsilon}x}u_{x}^2dx\bigg|+\alpha (V)\bigg].
\end{align}
\end{example}

\section{Proof of the Main Results}
\subsection{Proof of Theorem \ref{thm1}}
In this section, we prove Theorem \ref{thm1}. The proof follows in two steps. First, we show that the controller $\kappa_{c}^{*} := m\kappa_{c}(\phi,\beta/m^2,V)$ for $m\geq 2$ achieves asymptotic stability in the sense of inequality \eqref{exp_stab}. Then, we show that $v = \kappa_{c}^*$ is the unique asymptotically stabilizing minimizer of $\mathcal{J}_{c}$. 
\subsubsection{Step 1. The Feedback $\kappa_{c}^*$ is Asymptotically Stabilizing}
Let $v=\kappa_{c}^{*} := mk_{c}(\phi, \beta/m^2,V)$ for $m\geq 2$, where $\kappa_{c}$ is the feedback controller \eqref{feedback}. Note that we have 
\begin{align}
\beta v + v^3 =&~ m\beta \kappa_{c}(\phi, \beta/m^2,V)+ m^3 \kappa_{c}(\phi, \beta/m^2,V)^3 \nonumber \\
=&~ m^3\left(\frac{\beta}{m^2}\kappa_{c}(\phi, \beta/m^2,V) + \kappa_{c}(\phi, \beta/m^2,V)^3\right) \nonumber \\
=&~ -m^3\left(|\phi|+\frac{2\sqrt{3}}{9}\left|\frac{\beta}{m^2}\right|^{\frac{3}{2}}+\alpha(V)\right).
\end{align}
As a consequence, we obtain
\begin{align}
\dot{V} = \phi - m^3\left(|\phi|+\frac{2\sqrt{3}}{9}\left|\frac{\beta}{m^2}\right|^{\frac{3}{2}}+\alpha(V)\right), 
\end{align}
which implies, since $m\geq 2$, asymptotic stability in the sense of inequality \eqref{exp_stab}. 
\subsubsection{Step 2. Optimality}
We shall prove now that $\kappa_{c}^{*}$ is the unique minimzer of the cost $\mathcal{J}_{c}$. First, note that We have 
\begin{align}
\mathcal{J}_{c} =&~ -2m\int_{0}^{\infty} (\phi + \beta v + v^3 -\phi - \beta v - v^3)dt \nonumber \\
&~+\int_{0}^{\infty}(-2m(\phi - R_{c}^{-1})+m(m-2)R_{c}^{-1})dt \nonumber \\
&~+\int_{0}^{\infty}R_{c}(\beta + v^2)^2v^2dt.\label{jc1}
\end{align}
Using \eqref{depressed_cubic}, we can rewrite \eqref{jc1} as
\begin{align}
\mathcal{J}_{c} =&~ -2m\int_{0}^{\infty}dV +2m\int_{0}^{\infty}(\phi +\beta v + v^3)dt \nonumber \\
&~+\int_{0}^{\infty}(-2m(\phi - R_{c}^{-1})+m(m-2)R_{c}^{-1})dt \nonumber \\
&~+\int_{0}^{\infty}R_{c}(\beta + v^2)^2v^2dt \nonumber \\
=&~ -2m(V(\infty)-V(0))+2m\int_{0}^{\infty}(\phi +\beta v + v^3)dt \nonumber \\
&~+\int_{0}^{\infty}(-2m(\phi - R_{c}^{-1})+m(m-2)R_{c}^{-1})dt \nonumber \\
&~+\int_{0}^{\infty}R_{c}(\beta + v^2)^2v^2dt. \label{jc2} 
\end{align}
Since we are searching for the minimizer of $\mathcal{J}_{c}$ over asymptotically stabilizing controllers, we can set $V(\infty)=0$. Rearranging and simplifying the terms in \eqref{jc2}, we find 
\begin{align}
\mathcal{J}_{c} =&~ 2mV(0) +\int_{0}^{\infty}(m^2R_{c}^{-1}+2m(\beta v+ v^3))dt \nonumber \\
&~+\int_{0}^{\infty}R_{c}(\beta + v^2)^{2}v^2dt. \label{jc3}
\end{align}
Next, note that 
\begin{align}
&R_{c}(\beta v + v^3+mR_{c}^{-1})^2 = R_{c}((\beta + v^2)^2v^2+m^2R_{c}^{-2}\nonumber \\
&~+2m(\beta v+ v^3)R_{c}^{-1}) = R_{c}(\beta + v^2)^2v^2+m^2R_{c}^{-1}\nonumber \\
&~\quad \quad \quad \quad \quad \quad \quad \quad \quad \quad \quad \quad +2m(\beta v + v^3). \label{jc4}
\end{align}
Using \eqref{jc4}, we can rewrite \eqref{jc3} as 
\begin{align}
\mathcal{J}_{c} = 2mV(0)+\int_{0}^{\infty}(\beta v + v^3+mR_{c}^{-1})^2R_{c}dt. \label{jc5}
\end{align}
Using the facts that $- mR_{c}^{-1} = \beta \kappa_{c}^{*}+ (\kappa_{c}^{*})^3$, equation \eqref{jc5} becomes
\begin{align}
\mathcal{J}_{c} =&~ 2mV(0)+\int_{0}^{\infty}R_{c}(\beta v + v^3-\beta \kappa_{c}^{*}- (\kappa_{c}^{*})^3)^2dt.  
\end{align}
The feedback $v = \kappa_{c}^{*}$ is therefore optimal for the cost $\mathcal{J}_{c}$, and the minimum of $\mathcal{J}_{c}$ is $\mathcal{J}_{c}^{*} = 2mV(0)$. It remains to show that it is the unique minimizer. To do so, we show that the unique real root of the depressed cubic equation
\begin{align}
v^3+\beta v -\beta \kappa_{c}^{*}-(\kappa_{c}^{*})^3 = 0 \label{cubic_final}
\end{align}
is given by the Cardano root formula, that is, $v = \kappa_{c}^{*}$. This cubic equation admits a unique real root if its discriminant is positive, i.e. 
\begin{align}
\Delta := 4\beta ^3 + 27(\beta \kappa_{c}^{*}+(\kappa_{c}^{*})^3)^2 > 0. 
\end{align}
From the definition of $\kappa_{c}^{*}$ we have 
\begin{align}
\beta \kappa_{c}^{*}+(\kappa_{c}^{*})^3 = -m^3\left(|\phi|+\frac{2\sqrt{3}}{9}\left|\frac{\beta}{m^2}\right|^{\frac{3}{2}}+\alpha (V)\right). 
\end{align}
As a consequence, $\Delta$ becomes 
\begin{align}
\Delta =&~ 4\beta ^3 + 27m^6\left(|\phi|+\frac{2\sqrt{3}}{9}\left|\frac{\beta}{m^2}\right|^{\frac{3}{2}}+\alpha (V)\right)^2 \nonumber \\
=&~ 4\beta ^3 + 27m^6\frac{4}{27}\frac{|\beta|^3}{m^6}+27m^6\bigg(|\phi|^2+\alpha (V)^2\nonumber \\
&~ + 2|\phi|\alpha (V) + \frac{4\sqrt{3}}{9}\left|\frac{\beta}{m^2}\right|^{\frac{3}{2}}(|\phi|+\alpha (V))\bigg). \nonumber \\
\geq&~ \alpha (V)^2. 
\end{align}
Since $\alpha$ is a class $\mathcal{K}_{\infty}$ function, then $\Delta >0$ provided that $V\neq 0$, i.e. $u\neq 0$. In this case, the unique real root is $v=\kappa_{c}^{*}$. Moreover, if $u=0$, then $\beta = 0$ and the cubic equation \eqref{cubic_final} becomes $v^3=(\kappa_{c}^{*})^3$, which implies that $v=\kappa_{c}^{*}$. We conclude that the unique minimizer of $\mathcal{J}_{c}$ is $v=\kappa_{c}^{*}$. 

\subsection{Proof of Theorem \ref{thm2}}
This section is devoted to the proof of Theorem \ref{thm2}. We follow the same steps as in the proof of Theorem \ref{thm1}. Namely, we first show that the controllers $v = \kappa_{q}^{*} := m\kappa_{q}(\phi, \beta/m, V)$ for $m\geq 2$, and $v=\beta - \kappa_{q}^{*}$ achieve asymptotic stability in the sense of inequality \eqref{exp_stab}, then we show that $v=\kappa_{q}^{*}$ and $v=\beta -\kappa_{q}^{*}$ minimize the cost $\mathcal{J}_{q}$. 

\subsubsection{Step 1. The Feedback Laws $\kappa_{q}^{*}$ and $\beta - \kappa_{q}^{*}$ are Asymptotically Stabilizing}
 Let $v=\kappa_{q}^{*} := m\kappa_{q}$. Note that we have 
\begin{align}
\beta v -v^2 =&~ m\beta \kappa_{q}-m^2\kappa_{q}^2 = m^2\left(\frac{\beta}{m}\kappa_q -\kappa_{q}^2\right) =-m^2\theta (u).
\end{align}
Equation \eqref{quadratic_vdot} becomes 
\begin{align}
\dot{V} = \phi -m^2\left(|\phi|+\alpha (V)\right)\label{vdot_final}
\end{align}
which implies, since $m\geq 2$, asymptotic stability in the sense of inequality \eqref{exp_stab}.

Now, let $v = \beta - \kappa_{q}^{*}$. We have 
\begin{align}
\beta v - v^2 =&~ \beta (\beta - \kappa_{q}^{*})-(\beta - \kappa_{q}^{*})^2 \nonumber \\
=&~ \beta ^2 - \beta \kappa_{q}^{*} -\beta^2 - (\kappa_{q}^{*})^2+2\beta \kappa_{q}^{*} \nonumber \\
=&~ \beta \kappa_{q}^{*}-(\kappa_{q}^{*})^2 = -m^2 \theta (u),
\end{align}
which implies \eqref{vdot_final}, and therefore asymptotic stability in the sense of inequality \eqref{exp_stab}. 
\subsubsection{Step 2. Optimality}
We prove now that $v=\kappa_{q}^{*}$ and $v=\beta - \kappa_{q}^{*}$ minimize the cost $\mathcal{J}_{q}$. First, note that we have 
\begin{align}
\mathcal{J}_{q} =&~ -2m\int_{0}^{\infty} (\phi + \beta v - v^2 -\phi - \beta v + v^2)dt \nonumber \\
&~+\int_{0}^{\infty}(-2m(\phi - R_{q}^{-1})+m(m-2)R_{q}^{-1})dt \nonumber \\
&~+\int_{0}^{\infty}R_{q}(\beta - v)^2v^2dt.\label{jd1}
\end{align}
Using \eqref{quadratic_vdot} and the fact that $V(\infty)=0$ for asymptotically stabilizing controllers, we can rewrite \eqref{jd1} as 
\begin{align}
\mathcal{J}_{q} =&~ 2mV(0) +\int_{0}^{\infty}(m^2R_{q}^{-1}+2m(\beta v- v^2))dt \nonumber \\
&~+\int_{0}^{\infty}R_{q}(\beta - v)^{2}v^2dt. \label{jd2}
\end{align}
We note then the identity 
\begin{align}
R_{q}(\beta v -v^2+mR_{q}^{-1})^2 =&~ R_{q}(\beta -v)^2v^2+m^2R_{q}^{-1} \nonumber \\
&~+2m(\beta v - v^2). \label{quadratic_identity}
\end{align}
Using \eqref{quadratic_identity}, we can rewrite \eqref{jd2} as 
\begin{align}
\mathcal{J}_{q} = 2mV(0) +\int_{0}^{\infty}R_{q}(\beta v-v^2+mR_{q}^{-1})^2dt. \label{jd3} 
\end{align}
By observing that $-mR_{q}^{-1} = \beta \kappa_{q}^{*}-(\kappa_{q}^{*})^2$, equation \eqref{jd3} becomes 
\begin{align}
\mathcal{J}_{q} =&~ 2mV(0)+\int_{0}^{\infty}R_{q}(\beta v - v^2 - \beta \kappa_{q}^{*}+(\kappa_{q}^{*})^2)^2dt
\nonumber\\ 
=& ~
2mV(0)+\int_{0}^{\infty}R_{q}
(v-\kappa_{q}^{*})^2(\beta - \kappa_{q}^{*} - v)^2
dt
\end{align}
The cost is minimized with two distinct feedback laws,
$v=\kappa_{q}^{*}$ and $v=\beta -\kappa_{q}^{*}$. In both cases, the minimum of $\mathcal{J}_{q}$ is $\mathcal{J}_{q}^{*} = 2mV(0)$.  

\section{Switching between Inverse Optimal Controllers}\label{further_results}
An important property of the cost functional $\mathcal{J}_{q}$ in Theorem \ref{thm2} is that it does not admit a unique asymptotically stabilizing minimizer, to the contrary of the cost $\mathcal{J}_{c}$ in Theorem \ref{thm1} that is minimized only by the Cardano-Lyapunov controller. As a result, it is natural to wonder which of the two feedback laws, $v=\kappa_{q}^{*}$ or $v=\beta -\kappa_{q}^{*}$, one should choose. These two feedback laws lead to the same minimum for $\mathcal{J}_{q}$, namely $\mathcal{J}_{q}^{*} = 2mV(0)$. They are therefore not different in terms of optimality. To understand how $\kappa_{q}^{*}$ differs from $\beta - \kappa_{q}^{*}$, let us write their formula explicitly. We have 
\begin{align}
&\kappa_{q}^{*} = \frac{\beta}{2}+\frac{m}{2}\sqrt{\left(\frac{\beta}{m}\right)^2+4\theta }, \ \ \text{and} \\
&\beta -\kappa_{q}^{*} = \frac{\beta}{2}-\frac{m}{2}\sqrt{\left(\frac{\beta}{m}\right)^2+4\theta },
\end{align}
where $\theta = |\phi|+\alpha (V)$. At a given time instant $t$, it is better, in terms of input norm, to use $\kappa_{q}^{*}(u(t))$ instead of $\beta(u(t))-\kappa_{q}^{*}(u(t))$ if $\beta (u(t))\leq 0$, and vice-versa, it is better to use $\beta(u(t))-\kappa_{q}^{*}(u(t))$ instead of $\kappa_{q}^{*}(u(t))$ if $\beta (u(t))\geq 0$. This being said, we propose to use the feedback law $\kappa_{s}^{*} := \kappa_{q}^{*}$ if $\beta < 0$, and $\kappa_{s}^{*} = \beta -\kappa_{q}^{*}$ if $\beta >0$. This feedback law, that switches between the two inverse optimal controllers $\kappa_{q}^{*}$ and $\beta -\kappa_{q}^{*}$ is inverse optimal for the cost $\mathcal{J}_{q}$. It leads to the same minimum as before. Its advantage, over using only $\kappa_{q}^{*}$ or $\beta - \kappa_{q}^{*}$, is that it reduces the control effort. Strictly speaking, the switching feedback $\kappa_{s}^{*}$ is defined as follows
\begin{align}
v=& \kappa_{s}^{*}(u) :=    \arg_{\sigma\in\left\{\kappa_{q}^{*}, \beta - \kappa_{q}^{*}\right\}}\min\{|\sigma|\}
\nonumber\\[5pt]
= &
\left\{
\begin{array}{ll}
 \displaystyle
- \sign(\beta)\frac{2m^2 \theta }{|\beta| +\sqrt{\beta^2+4m^2\theta } }\,,
     &  \beta \neq 0 
      \\[15pt]
 \pm m\sqrt{\theta}\,, & \beta =0.
\end{array}
\right.
\label{ks}
\end{align}
The discontinuity of $\kappa_s^*$ occurs along the axis $\beta = 0$. It is worth noting that the cost $\mathcal{J}_{q}$ does not distinguish between the controllers $\kappa_{q}^{*}$, $\beta - \kappa_{q}^{*}$ and $\kappa_{s}^{*}$, because they all lead to the same value for $(\beta -v)^2v^2$. It also means that, although the closed-loop solutions are different, the differential equation on $\dot{V}$ remains the same, namely \eqref{diff_eq_V}. This observation is specific to the fact that the actuated part of $\dot{V}$ is a quadratic polynomial admitting two real roots that are distinct when $\beta \neq 0$, rather than an affine function in the control.
\begin{remark}
Note that this idea of switching from one real root of a quadratic equation to the other root, depending on the sign of the function $\beta$ that multiplies $v$, appears e.g. in \cite[Equation 27]{espitia_lyapunov} for the special case of linear diffusion-reaction systems with right-end Dirichlet actuation. This approach can also be employed when the derivative of the CLF has a cubic dependence in the control input, as shown in \cite[Lemma 6]{KS1}, \cite[Lemma 3]{ACC23-KS} and \cite[Equations 6 and 7]{auto_KS} for the Kuramoto-Sivashinsky equation.
\end{remark}
\section{Conclusion}
We proved in this paper the inverse optimality of the Cardano-Lyapunov controller when the derivative of the CLF has a depressed cubic dependence in the control input, and we constructed a cost functional that is minimized by two distinct controllers for the case where the derivative of the CLF is quadratic in the control. We have also shown, in the second case, how to switch from one inverse optimal controller to the other to reduce the control effort. In future work, we would like to extend our approach to the problem of inverse optimal safety filter design for systems that are not affine in the control.
 
\bibliography{biblio}
\bibliographystyle{IEEEtranS}     

\end{document}